\documentclass[aps,twocolumn,superscriptaddress,floatfix,pra]{revtex4}

\usepackage{amsmath,amssymb,graphicx}
\usepackage{color}

\graphicspath{{figures/}}

\begin{document}

\date{\today}

\title{\textbf{On the Bifurcation and Stability of Single and Multiple Vortex Rings
in Three-Dimensional Bose-Einstein Condensates}}

\author{Russell Bisset}
\email{rnbisset@gmail.com}
\affiliation{Center for Nonlinear Studies and Theoretical Division, Los Alamos
National Laboratory, Los Alamos, NM 87544}

\author{Wenlong Wang}
\affiliation{Department of Physics, University of Massachusetts,
Amherst, Massachusetts 01003 USA}

\author{C. Ticknor}
\affiliation{Theoretical Division, Los Alamos
National Laboratory, Los Alamos, NM 87544}

\author{R. Carretero-Gonz{\'a}lez}
\affiliation{Nonlinear Dynamical Systems
Group,\footnote{\texttt{URL}: http://nlds.sdsu.edu}
Computational Sciences Research Center, and
Department of Mathematics and Statistics,
San Diego State University, San Diego, California 92182-7720, USA}

\author{D.J. Frantzeskakis}
\affiliation{Department of Physics, University of Athens,
Panepistimiopolis, Zografos, Athens 15784, Greece}

\author{L.A. Collins}
\affiliation{Theoretical Division, Los Alamos
National Laboratory, Los Alamos, NM 87544}

\author{P.G. Kevrekidis}
\email{kevrekid@math.umass.edu}
\affiliation{Department of Mathematics and Statistics, University of Massachusetts,
Amherst, Massachusetts 01003-4515 USA}

\affiliation{Center for Nonlinear Studies and Theoretical Division, Los Alamos
National Laboratory, Los Alamos, NM 87544}

\begin{abstract}
In the present work, we investigate how single- and multi-vortex-ring
states can emerge from a planar dark soliton in three-dimensional (3D)
Bose-Einstein condensates (confined in isotropic or anisotropic traps)
through bifurcations. We characterize such bifurcations
quantitatively using a Galerkin-type approach, and find good qualitative and
quantitative agreement with our Bogoliubov-de Gennes (BdG) analysis.
We also systematically characterize the BdG spectrum
of the dark solitons, using perturbation theory, and obtain a quantitative match with our 3D
BdG numerical calculations.
We then turn our attention to the emergence of
single- and multi-vortex-ring states. We systematically capture these
as stationary states of the system and quantify their BdG spectra numerically.
We find that although the vortex ring may be unstable when
bifurcating, its instabilities weaken and may even eventually disappear,
for sufficiently large chemical potentials and suitable trap settings.
For instance, we demonstrate the stability of the vortex ring for an
isotropic trap in the large chemical potential regime.
\end{abstract}

\maketitle

\section{Introduction}

Over the last two decades, there has been an intense interest in
nonlinear matter waves
in the context of atomic Bose-Einstein condensates (BECs)~\cite{book1,book2,emergent}.
In the three-dimensional (3D) setting, arguably, one of the most prototypical excitations
that arise are vortex rings (VRs). Such toroidal-shaped vortices, or rings
of vorticity, have been predicted theoretically
and observed experimentally in numerous studies; see, e.g.,
the relevant reviews~\cite{emergent,komineas_rev,book_new}.
In addition to their relevance in superfluids, VRs are of intense interest
in other areas, e.g., in fluid mechanics~\cite{saffman,Pismen1999}.
Superfluid VRs have been observed experimentally in
helium \cite{donnelly,Rayfield64, Gamota73}, far before
their emergence in atomic BECs of dilute alkali gases.
In the atomic BEC setting, these states have
arisen in a variety of ways. One such example
is through the bending of a 3D vortex line (i.e., a line of vorticity) so that it closes on itself~\cite{VR:CrowInstab}.
Vortex rings have also been experimentally realized through
the decay of planar dark solitons in two-component BECs~\cite{Anderson01},
by density engineering~\cite{Shomroni09} (related to earlier
theoretical proposals of Refs.~\cite{Ruostekoski01,Ruostekoski05}),
and in the evolution of colliding symmetric defects~\cite{Ginsberg05}.
They have also been detected through their
unusual collisional outcomes of structures that
may appear as dark solitons in cigar-shaped traps~\cite{sengstock}.

Numerical studies have also theoretically explored potential methods to
generate VRs. Some of these examples involve the
flow past an obstacle~\cite{Jackson99,Rodrigues09},
Bloch oscillations in an optical trap~\cite{Scott04},
the collapse of bubbles~\cite{Berloff04a},
the instability of two-dimensional (2D) rarefaction pulses~\cite{Berloff02},
the flow past a positive ion~\cite{Berloff00,Berloff00a} or an electron bubble~\cite{Berloff01},
the Crow instability of two vortex pairs~\cite{Berloff01a} and
the collisions of multiple BECs~\cite{Carretero08,Carretero08a}.
VR cores have an intrinsic velocity~\cite{Roberts71}
(a feature that can be understood by considering their cross-section
resemblance to a vortex dipole which is well-known to travel at a
constant speed~\cite{saffman,Pismen1999}). However, this intrinsic velocity can be
counterbalanced by the presence of an external trap, which is commonplace in
atomic BECs~\cite{VR-BEC-STRUCTGOOD}.
This, in turn, creates the possibility for the existence of stationary VRs, and, in fact, even of multi-rings, as discussed earlier,
e.g., in Ref.~\cite{ionut}.

The aim of the present work is to revisit the study of VRs,
but from a different perspective to the above works, namely we focus
of their bifurcation and emergence in the weakly nonlinear limit.
In this sense, our study is partially connected to the recent
work of Ref.~\cite{pantof}, where a somewhat similar approach was adopted. 
Nevertheless, their work focused on a setting where the trapping was only in two out
of three spatial directions; another fundamental difference with
respect to Ref.~\cite{pantof}
is their absence of stability information
in the relevant setup. On the other hand, our study is also inspired by the pioneering
work of Ref.~\cite{feder}, which, in turn, motivated the experimental results
of Ref.~\cite{Anderson01}. In Ref.~\cite{feder}, the spectrum of a dark
soliton was examined (although not in its full detail, as we will
discuss below) and some of
the key instability mechanisms associated with it were elucidated.
We 
are also motivated by the analysis presented in
Refs.~\cite{middelphysd,middel2},
which used a somewhat similar approach to that used herein
to explore bifurcations from a two-dimensional dark soliton
to vortex-dipoles and, more generally, multi-vortex structures.

Our fundamental starting point will be the realization that a VR
emerges from a suitable combination of two distinct states with
a relative phase (of $\pi/2$ between them), namely a
planar dark soliton (say, along the $z$-direction) and of a
ring dark soliton (RDS)
along the $(x,y)$ plane, another state studied extensively in earlier work;
see, e.g., Refs.~\cite{rds2003,herring,kamch}.
Should the planar dark soliton have a lower energy than the RDS,
then the bifurcation happens from the former; if the order of
energies is reversed, then the bifurcation 
occurs from the RDS.
If, however, the energies are equal (which, as we will see, happens
for anisotropic traps, i.e., when the trapping
frequencies along the $z$- and radial $r$- 
directions satisfy $\omega_z=2 \omega_r$),
the VR state already emerges at the linear (noninteracting) limit. The presence
of two principal modes in the coherent structure enables us to adapt a
Galerkin-type, two-mode methodology, originally developed for double-well potentials;
see, e.g., relevant details in Ref.~\cite{theochar}, based on the fundamental
earlier work of Ref.~\cite{smerzi}. This two-mode approach allows us to provide a
quantitative estimate, which turns out to not only qualitatively but
also quantitatively predict the bifurcation/emergence point of the single VR.
We use a similar approach to generalize to
more complex states such as the double VR (arising due to
the combination of a double-soliton/second excited state along the $z$-axis,
with a RDS along the $(x,y)$ plane), and we further
extend this approach to single and double vortex lines.

Since the planar dark soliton may spawn the VR, in the appropriate regime, we expand upon the work of Ref.~\cite{feder} by capturing all its modes using the Bogoliubov-de Gennes (BdG) analysis near the linear (small-interaction) limit. 
%
%
This limit enables a perturbative treatment of the relevant modes,
both in the isotropic and in the anisotropic case, clearly allowing us to identify
neutral modes, modes responsible for instabilities and
also the bifurcations of novel states such as the VR,
and the single and double solitonic vortex.
We provide, both, analytical estimates and full-numerical calculations
for the relevant eigenvalues, and provide connections with the analytical
two mode theory discussed above.
Once the relevant bifurcations arise, especially those leading to the single and double VRs
of focal interest herein, we turn our attention to their spectra to quantify instability and the potential stabilization
mechanisms.
This leads us to conclude that these states tend to be weakly unstable in the small-interaction limit, but may ---in principle--- be stabilized in the large-interaction setting.

It is relevant to mention that this problem remains of particular interest,
not only for theoretical but also for recent experimental studies.
A notable example is the very recent work of Ref.~\cite{zwierlein}, where
the tomographic imaging of a superfluid Fermi gas of $^6$Li atoms
enabled the observation of the snaking instability of the planar
dark soliton into a VR (and subsequently into a vortex line).
Connecting state-of-the-art observations such as these to the understanding of stationary-state stability offered here should furnish a complete picture of the underlying complex dynamics of the system.

Our presentation is structured as follows. Section \ref{sec:analytical}
contains the
different aspects of our analytical results; first, the bifurcation
analysis (of single and double VRs) using the Galerkin method; and second, the spectral
analysis, for the small-interaction limit, of the dark solitons
which we connect to the above bifurcation structure. In Sec.~\ref{sec:numerics},
we provide detailed numerical computations of the existence and stability
of dark solitons, and subsequently VRs and multi-VRs,
starting from the small-interaction limit. We provide comparisons between our numerical and
analytical predictions. Finally, in Sec.~\ref{sec:conclu}, we summarize our findings
and present both, our conclusions, as well as some interesting directions
for future studies.

\section{Analytical Considerations}
\label{sec:analytical}

In this work we utilize the 3D Gross-Pitaevskii
equation (GPE), expressed in the following dimensionless 
form~\cite{emergent}:
\begin{equation}
i u_t=-\frac{1}{2} \nabla^2 u +V(r) u +| u |^2 u.
\label{GPE}
\end{equation}
Here, $u$ 
is the macroscopic wavefunction of the 3D BEC
near zero temperature, while the potential assumes the prototypical parabolic form, namely:
\begin{eqnarray}
V(x,y,z)= \frac{1}{2} \omega_r^2 r^2 + \frac{1}{2} \omega_z^2 z^2.
\label{pote}
\end{eqnarray}
The parameters $\omega_r$ and $\omega_z$ represent the trapping strengths along the
$(x,y)$ plane and $z$ direction, respectively, with 
the spherically symmetric (isotropic) case corresponding to $\omega_r=\omega_z$.

\subsection{Vortex Ring Bifurcation Near the Linear Limit}


In the linear limit, Eq.~(\ref{GPE}) reduces to the
quantum harmonic oscillator (QHO), with energy spectrum: 
\begin{eqnarray}
E_{n,m,k}= \omega_r \left( n + m + 1 \right) +
\omega_z \left( k + 1/2 \right),
\label{linear}
\end{eqnarray}
where $n$, $m$ and $k$ are the non-negative integers indexing the
corresponding eigestate.
Let us now discuss how to construct a VR starting from
the considered linear limit. Intriguingly, utilizing the results of Ref.~\cite{ionut},
this is possible in the following way: 
in that work it was found that for the anisotropic trap with $\omega_r=1$ and $\omega_z=2/k$
the energy of a second radial excited state ($m$,$n$,$k^\prime$) with $m+n=2$ and $k^\prime=0$ (i.e.~$E=3+1/k$) coincides with that of the $k$-th excited state along
the $z$-axis $(0,0,k)$ (also equal to $1+ (2/k) (k+1/2)=3+1/k$).
For such anisotropic traps one can construct stationary states with $k$-parallel-VRs
\begin{eqnarray}
u(x,y,z) \propto \frac{|2,0,0\rangle + |0,2,0\rangle}{\sqrt{2}} + i |0,0,k \rangle
\label{VR} ,
\end{eqnarray}
{\it even at the linear limit}.
The RDS nature of the real part combined with the $k$ oscillations of the imaginary part along the $z$-direction constitutes parallel VRs with alternating vorticity.
Remarkably, this is a very natural 3D generalization of the 2D setting considered in Ref.~\cite{middelphysd}. In the above expression of Eq.~(\ref{VR}),
we have used the notation
\begin{eqnarray}
|n,m,k \rangle&=&
H_n ( \sqrt{\omega_r} x)
H_m ( \sqrt{\omega_r} y)
H_k ( \sqrt{\omega_z} z)
\nonumber \\
&\times&\exp[-(\omega_r (x^2+y^2) +\omega_z^2 z^2)/2],
\end{eqnarray}
corresponding to the eigenmode of energy $E_{n,m,k}$ of the QHO
where $H$ represents the Hermite polynomials.

Generalizing the approach of Ref.~\cite{middelphysd}
by considering anisotropic trap strengths $\omega_r$
and $\omega_z$ (including the isotropic one of $\omega_r=\omega_z$),
the energies of the RDS and the $k$-th solitonic state will, respectively be,
$$
E_{\rm RDS}=3 \omega_r + \frac{\omega_z}{2},
\quad {\rm and} \quad
E_{\rm sol}=\omega_r +(k+\frac{1}{2}) \omega_z.
$$
Hence, for $k \omega_z < 2 \omega_r$,
the state of lower energy will be the $k$-th soliton
from which the $k$-th VR will bifurcate,
while for $k \omega_z > 2 \omega_r$, the lower energy will belong to the
RDS, with the bifurcation occurring from that state.

\subsubsection{Single vortex ring, $\omega_z<2\omega_r$}

Following the bifurcation phenomenology of Ref.~\cite{middelphysd}
(see also Ref.~\cite{middel2}), we first consider the limit $\omega_z < 2 \omega_r$ which encompasses the isotropic trap, and proceed as follows.
Take the $k$-th soliton along the $z$-axis, focusing for now on $k=1$, and assign this as mode $u_1$ with energy $E_1$.
Similarly, assign the RDS as mode $u_2$ with energy $E_2$.
If a novel state (in this case, the single VR) bifurcates from these two states when their phase difference is $\pi/2$, then a general two-mode analysis~\cite{theochar} predicts that the number
of atoms $N=\int |u|^2 dx dy dz$ at the bifurcation critical point will be given by
\begin{eqnarray}
N_{\rm cr}= \frac{E_1 -E_2}{I_{12} - I_{11}},
\label{predict}
\end{eqnarray}
where for the above special case of the single VR we have
$E_1-E_2= \omega_z-2 \omega_r$, while
$I_{11}=\int |u_1|^4 dx dy dz$ and $I_{12}= \int |u_1|^2 |u_2|^2 dx dy dz$
are overlap integrals.
Specifically, for the VR, the dark soliton state from which the bifurcation arises is $u_1=|0,0,1 \rangle$, while the RDS is $u_2=\frac{|2,0,0\rangle + |0,2,0\rangle}{\sqrt{2}}$.

Furthermore, the general two-mode theory also predicts the chemical potential value at which the bifurcation
will occur, namely
\begin{equation}
\mu_{\rm cr}=E_1 + I_{11} N_{\rm cr},
\end{equation}
where $N_{\rm cr}$ is given by 
Eq.~(\ref{predict}).

Interestingly, the above integrals can be computed analytically.
In particular, for the case of the single VR with
$k=1$, we find that
\begin{eqnarray}
I_{11} = \frac{3 \omega_r \sqrt{\omega_z}}{8 \sqrt{2} \pi^{3/2}}=3 I_{12}.
\label{overlap}
\end{eqnarray}
As a result, the explicit prediction for the bifurcation of the VR is that:
\begin{eqnarray}
N_{\rm cr}^{\rm (VR,1)} &=& \frac{4 \sqrt{2} \pi^{3/2}  (2 \omega_r-\omega_z)}{\omega_r \sqrt{\omega_z}},
\label{pred1}
\\
\mu_{\rm cr}^{\rm (VR,1)} &=& 4 \omega_r.
\label{pred2}
\end{eqnarray}
This provides us with an explicit prediction for the bifurcation
of the first VR.

\subsubsection{Single vortex ring, $\omega_z>2\omega_r$}

For completeness, we now consider the case of $\omega_z>2 \omega_r$, where the VR
now bifurcates from the RDS and the subscripts of Eq.~(\ref{predict})
must be exchanged 1$\leftrightarrow$2.
For this case, $E_2-E_1= 2 \omega_r - \omega_z$ and $I_{22}=\int |u_2|^4 dx dy dz
=2 I_{12}$, which leads to the bifurcation point:
\begin{eqnarray}
N_{\rm cr}^{\rm (VR,2)} &=& \frac{8 \sqrt{2} \pi^{3/2}  (\omega_z - 2 \omega_r)}{\omega_r \sqrt{\omega_z}}
\label{pred1a}
\\
\mu_{\rm cr}^{\rm (VR,2)} &=& \frac{5}{2} \omega_z - \omega_r .
\label{pred2a}
\end{eqnarray}
The superscripts in Eqs.~(\ref{pred1}) and (\ref{pred2}) versus those of Eqs.~(\ref{pred1a})
and (\ref{pred2a}) are used to illustrate which state the VR bifurcates from,
1 is for the dark soliton and 2 represents the RDS.

\subsubsection{Double vortex rings, $\omega_z < \omega_r$}

Similarly, for the double-vortex-ring (2VR) state
$E_1-E_2=2 (\omega_z-\omega_r)$, while the overlap integrals
are given by
\begin{eqnarray}
I_{11}=\frac{41 \omega_r \sqrt{\omega_z}}{128 \sqrt{2} \pi^{3/2} }=\frac{41}{12}
I_{12}.
\label{overlap2}
\end{eqnarray}
Hence, the corresponding prediction for the bifurcation point yields
\begin{eqnarray}
N_{\rm cr}^{\rm (2VR,1)} &=& \frac{256 \sqrt{2} \pi^{3/2}  (\omega_r-\omega_z)}{29 \omega_r \sqrt{\omega_z}},
\label{pred3}
\\
\mu_{\rm cr}^{\rm (2VR,1)} &=& \frac{222 \omega_r - 19 \omega_z}{58}.
\label{pred4}
\end{eqnarray}
These predictions are valid for $\omega_z \leq \omega_r$.
In this case, the 2VR will
bifurcate from the two-dark-soliton state of
the form $u_1=|0,0,2 \rangle$, while the higher energy
state in the two-mode analysis will again be the RDS.

\subsubsection{Double vortex rings, $\omega_z > \omega_r$}

On the other hand, for $\omega_z>\omega_r$ the prediction needs to be suitably modified.
Since the 2VR now bifurcates from the RDS we again exchange the subscripts in Eq.~(\ref{predict}).
Using $I_{22}=(41 \omega_r \sqrt{\omega_z})/(128 \sqrt{2} \pi^{3/2})$
(for $u_2=|0,0,2 \rangle$) we finally find that:
\begin{eqnarray}
N_{\rm cr}^{\rm (2VR,2)} &=& \frac{64 \sqrt{2} \pi^{3/2}  (\omega_z-\omega_r)}{5 \omega_r \sqrt{\omega_z}},
\label{pred3a}
\\
\mu_{\rm cr}^{\rm (2VR,2)} &=& \frac{37 \omega_z - 2 \omega_r}{10}.
\label{pred4a}
\end{eqnarray}

An interesting feature is the following.
For $\omega_z < 2 \omega_r$, the single VR bifurcates from the one-dark soliton ($k=1$) and
for $\omega_z < \omega_r$ the 2VR bifurcates from
the two-dark-soliton state ($k=2$).
In the ``intermediate'' case, $\omega_r < \omega_z < 2 \omega_r$,
the 2VR state bifurcates from the RDS,
while the single VR bifurcates from the one-dark soliton. Finally, for
$2 \omega_r < \omega_z$, both the single VR and the
2VR states bifurcate from the RDS.
Hence, as expected, $\mu_{\rm cr}^{\rm (2VR,2)} > \mu_{\rm cr}^{\rm (VR,2)}$, i.e.~first the lower-energy single VR bifurcates, followed by the 2VR state.

We remark in passing that, in line with the special-case
results of Ref.~\cite{ionut}, the single VR bifurcates from the
linear limit ($N \rightarrow 0$)
in the anisotropic case of $\omega_z=2 \omega_r$,
as per Eq.~(\ref{pred1}), while the 2VR bifurcates from the
linear limit, precisely for the isotropic case of $\omega_z=\omega_r$,
as per Eq.~(\ref{pred3}).
One can similarly generalize these types of bifurcation considerations
for all higher-order rings, providing an explicit set of predictions
for their emergence in the vicinity of the linear limit.
This bifurcation and stability analysis naturally
explains why the decay of dark solitonic states yields the corresponding $k$-VR
states, not only in numerical simulations~\cite{feder}, but also in
experiments~\cite{Anderson01}.

\subsubsection{Solitonic vortices}

While the emphasis here is on VRs, partly to illustrate that the different states obtained in Ref.~\cite{pantof} can be identified with the techniques presented herein, we also provide other special cases, namely the single- and double-solitonic vortex states.

For the case $\omega_z < \omega_r$, following the same technique as above but with $u_2= |1,0,0 \rangle$, we obtain:
\begin{eqnarray}
N_{\rm cr}^{\rm (1SV,1)} &=& \frac{4 \sqrt{2} \pi^{3/2}  (\omega_r-\omega_z)}{\omega_r \sqrt{\omega_z}},
\label{pred5}
\\
\mu_{\rm cr}^{\rm (1SV,1)} &=& \frac{5 \omega_r}{2},
\label{pred6}
\end{eqnarray}
for the single solitonic vortex (bifurcating from the
dark soliton $u_1=|0,0,1\rangle$).
For the double solitonic vortex, again bifurcating from $u_1=|0,0,1\rangle$ but with $u_2=(|2,0,0 \rangle - |0,2,0 \rangle)/\sqrt{2}$, one finds the bifurcation point:
\begin{eqnarray}
N_{\rm cr}^{\rm (2SV,1)} &=& \frac{4 \sqrt{2} \pi^{3/2}  (2 \omega_r-\omega_z)}{\omega_r \sqrt{\omega_z}},
\label{pred7}
\\
\mu_{\rm cr}^{\rm (2SV,1)} &=& 4 \omega_r,
\label{pred8}
\end{eqnarray}
for $\omega_z < 2 \omega_r$.
The reverse trap anisotropies can similarly be explored.

\subsubsection{Comparison to other work}

Finally, and before proceeding with the stability analyses
of the dark soliton and the resulting VR states that
are the principal focus of the present work, let us compare our work to Ref.~\cite{pantof}.
Our existence results bear significant resemblance to those of Ref.~\cite{pantof},
although there are also nontrivial differences.
For instance, in Ref.~\cite{pantof} the focus was on bifurcations from the dark soliton (called the kink therein) state, whereas we especially
focus on bifurcations from various states that yield VR-like
solutions (single and double VRs).
In Ref.~\cite{pantof}, the $z$-direction is presumed
to be homogeneous (i.e., untrapped), while here we consider the presence of a trap.
For this reason, multi-VR states (such as those explored
above) are {\it not} possible in the framework of Ref.~\cite{pantof}.
In fact, the 2VR state in that work is one in which both rings are in the same plane, while 
here we explore multiple {\it non-coplanar} rings in a stationary state, rendered
possible by the additional trapping along the $z$-direction.
On a technical level too, the methodology of the
reduction to a quasi-linear equation on the plane with
an effective potential used in Ref.~\cite{pantof}, while
ingenious, differs substantially from the two-mode
approach utilized herein.
Finally, and perhaps most importantly, a principal focus that stems from our
existence findings is that, we also study the stability of the obtained solutions 
(see details in the following section). In Ref.~\cite{pantof}, on the other hand,
this is deferred to future studies.

\subsection{Stability Analysis Near the Linear Limit}

We now move to the stability analysis in the vicinity of the linear
limit. Using a Taylor expansion of the solution $u= \sqrt{\epsilon} u_0 +
\epsilon^{3/2} u_1 + \dots$ and $\mu=\mu_0 + \epsilon \mu_1 + \dots$
in Eq.~(\ref{GPE}), where $(\mu_0,u_0)$ correspond, respectively, to the
eigenvalue and eigenfunction of a state
at the linear limit, we find at O$(\epsilon)$ the solvability condition:
\begin{eqnarray}
\mu_1= \int |u_0|^4 dx dy dz.
\label{solve}
\end{eqnarray}
This, in turn, allows us to specify $\epsilon= (\mu - \mu_0)/\mu_1$
(to leading order). Then, the spectral stability, as discussed
in Ref.~\cite{feder}, amounts to solving the BdG eigenvalue problem
$({\cal H}_0 + \epsilon {\cal H}_1) U= \omega U$, with
\begin{eqnarray}
U &= \left(\begin{array}{c}
\mathcal{U}\\
\mathcal{V}\\
\end{array} \right) , \label{Eq:uv} \\
{\cal H}_0 &=\left( \begin{array}{cc}
{\cal L} & 0 \\
0  & -{\cal L} \\
    \end{array} \right),
\end{eqnarray}
where ${\cal L}=-(1/2) \nabla^2  +V(r) - \mu_0$, while
\begin{eqnarray}
{\cal H}_1= \left( \begin{array}{cc}
2 |u_0|^2 - \mu_1 & u_0^2 \\
- (u_0^2)^{\star}  & \mu_1 - 2 |u_0|^2 \\
    \end{array} \right),
\end{eqnarray}
where the star denotes complex conjugate.
Here, we denote $\omega$ as the eigenfrequency of a given eigenstate $U$ with
eigenvalue $\lambda=i \omega$. The presence of a nonzero
imaginary part of $\omega$ or, equivalently,
of a nonzero real part of $\lambda$
in our Hamiltonian system denotes the presence of
a dynamical instability.

\subsubsection{The dark soliton}

From the above formulation, it is straightforward to analyze
the stability in the case of $\epsilon \rightarrow 0$.
There, it is evident that the linearization spectrum
consists of the diagonal contributions of
${\cal H}_0$ which consist of the spectrum of ${\cal L}$
and of its opposite.
Up to now, we have kept our exposition as general as possible, but
from here on, we will focus on a specific state in order to
showcase the relevant ideas more concretely. As our workhorse,
we will use the dark soliton state $u_0= |0,0,1 \rangle$, in
a spirit similar to that of Ref.~\cite{feder}, but with a particular view
towards the bifurcation of the VR state (as well as others such as the
2VR and the single and double solitonic vortex).
Since ${\cal L}$ consists of the quantum
harmonic oscillator, the spectrum for a dark soliton along the $z$-axis,  when subtracting $\mu_0= \omega_r + 3\omega_z/2$, will be (see Eq.~(\ref{linear}))
\begin{eqnarray}
\omega= \omega_r \left( n + m \right) + \omega_z (k-1).
\label{linear2}
\end{eqnarray}

Recalling that $n$, $m$, and $k$ are arbitrary non-negative integers, we
now proceed to explore the relevant states and their multiplicities.
The mode with $n=m=k=0$ corresponds to the
{\it negative energy}~\cite{book2,feder},
or negative Krein signature~\cite{kks},
mode. Such a mode, when resonant with another positive energy
(or Krein signature) one, will give rise to complex eigenvalue
quartets, while collisions of modes of the same energy will be
inconsequential towards changing the stability properties of
the solution. The dark soliton state (with vanishing density in the $z=0$
plane) has a neutral mode, corresponding to $(n,m,k)=(0,0,1)$,
which has a wavefunction identical to the solution itself.
This mode corresponds to the phase or gauge (U$(1)$) invariance
of Eq.~(\ref{GPE}). There are three additional Kohn modes corresponding
to symmetry which are also left invariant. Specifically, $(n,m,k)=(0,0,2)$
corresponds to dipolar oscillations along the $z$-axis with frequency $\omega=\omega_z$,
while the modes $(n,m,k)=(1,0,1)$ and $(n,m,k)=(0,1,1)$ pertain
to dipolar oscillations along the $(x,y)$-plane with frequency $\omega=\omega_r$.
To complete our discussion of modes with $n+m+k \leq 2$, we
need to account for five more modes. There are two degenerate
modes (due to the radial invariance of the trap in the plane)
of $(n,m,k)=(1,0,0)$ and $(n,m,k)=(0,1,0)$, with
frequency $\omega=\omega_r-\omega_z$ and
three degenerate ones, $(n,m,k)=(2,0,0)$, $(n,m,k)=(0,2,0)$
and $(n,m,k)=(1,1,0)$, all of which have frequency
$\omega=2 \omega_r-\omega_z$ in the linear limit.

\subsubsection{The dark soliton, $\omega_r=\omega_z$}

The key question that subsequently arises is that
of the fate of the eigenvalues described above, as $\epsilon$ becomes finite,
i.e., as we depart from the linear limit.
To follow these eigenvalues, and given that the different degeneracies
also hinge on the specific values of
$\omega_r$ and $\omega_z$, we will use as our benchmark
case the isotropic scenario of $\omega_r = \omega_z = 1$,
where the choice of unity is made without loss of generality.
In this case, the zero eigenvalue has a multiplicity of three.
The gauge-invariance mode must remain at zero, and so too must the two modes (1,0,0) and (0,1,0) due to the spherical invariance of the dark soliton solution, but only when the trap is isotropic.


We now move to the consideration of the principal $7$ modes (recall that this is
really 7 pairs) at $\omega=\pm 1$. Out of these, the 3 dipolar modes
will remain invariant. Then, however, 4 modes are subject
to deviations, as soon as we depart from the linear limit.
It turns out that the insightful work of Ref.~\cite{feder} has
already computed one sub-manifold associated with such
a bifurcation. In that work, the authors recognized that the eigenvector
associated with the RDS
$U_1=(\frac{|2,0,0\rangle + |0,2,0\rangle}{\sqrt{2}},0)^T$ [see Eq.~(\ref{Eq:uv})] and
the anomalous mode $U_2=(0,|0,0,0\rangle)^T$ become {\it resonant},
and identified the deviations
by using degenerate perturbation theory, i.e., constructing the
matrix ${\cal M}$ with
\begin{eqnarray}
{\cal M}_{ij}= \langle U_i | {\cal H}_1 | U_j \rangle,
\label{degen}
\end{eqnarray}
and identifying its eigenvalues for $i,j=1,2$
for the above eigenvectors.
That calculation, adapted to the present setting, can be rewritten as
\begin{eqnarray}
\omega=1 - \frac{\left(3 \pm i \sqrt{7} \right)}{12} \left(\mu- \frac{5}{2}
\right)
\label{eig1}
\end{eqnarray}
(cf.~Eq.~(20) in Ref.~\cite{feder}).
It is important to note that the near-linear prediction is that the
dark soliton should be {\it immediately unstable} due to this resonant
interaction, via an oscillatory instability and a quartet of corresponding
eigenvalues. This is a general feature that dark soliton (and
multi-soliton) states possess near the linear limit due to the degeneracy
of their anomalous modes; cf.~the work of Ref.~\cite{coles}. However, in that work these instability ``bubbles'' were shown to terminate at some finite
value of $\mu$, hence it is of interest to explore whether a similar
feature arises here, a question that we will address below numerically.

On the other hand, in the work of Ref.~\cite{feder}, an additional
(and, as we will see, important) sub-manifold of eigenvectors was not
considered, namely that of
$U_1=(\frac{|2,0,0\rangle - |0,2,0\rangle}{\sqrt{2}},0)^T$
and of $U_2=(|1,1,0\rangle,0)^T$. For this subspace, we find
that the corresponding deviation of the eigenfrequency from the linear
limit ---again, obtained via the degenerate perturbation theory
of Eq.~(\ref{degen})--- yields:
\begin{eqnarray}
\omega=1 - \frac{2}{3} \left(\mu - \frac{5}{2} \right).
\label{eig2}
\end{eqnarray}

Given the decreasing trend of both of these eigenvalues, it is natural
to expect that at some finite value of $\mu$, they will hit the
origin of the spectral plane. As a preamble to our numerical computations
of the next section, it is then relevant to consider what the outcome
of such a collision will be. Connecting these findings
with our bifurcation theory results of the previous
subsection, we appreciate that these collisions should be, in
fact, what leads to the corresponding bifurcations and destabilizations
of the dark soliton along the $z=0$ plane. In particular,
the mode $U_1=(\frac{|2,0,0\rangle + |0,2,0\rangle}{\sqrt{2}},0)^T$
effectively corresponds to the RDS. Its
``collision'' with the origin and subsequent destabilization
of the dark soliton along the $z=0$ plane suggests that
beyond this threshold the planar dark soliton and RDS mesh,
which, as we have discussed before, produces the single VR
state. In the same spirit, we can see that the
collision of $U_1=(\frac{|2,0,0\rangle - |0,2,0\rangle}{\sqrt{2}},0)^T$
with the origin will produce a meshing with the planar dark soliton and lead to the bifurcation of
the two solitonic vortex state, discussed in the previous subsection.
In fact, assuming the prediction of Eq.~(\ref{eig2}) to be useful
beyond its realm of validity (and until $\omega=0$) yields
a critical point for the relevant bifurcation at $\mu=4$, which
coincides with the prediction of $\mu^{\rm (2SV,1)}$ of Eq.~(\ref{pred8}).
We will numerically examine the validity of these predictions
in the next section.

\subsubsection{The dark soliton, $\omega_r=2\omega_z$}

In principle, the approach adopted above (and the associated eigenvalue
count and deviations from the linear limit) can be used for any state bifurcating
from the linear limit. However, obviously, the more complicated
the original state, the more difficult it becomes to account for
all the relevant eigenvalues. Since our focus here is on
VRs and their emergence, we will also give an additional
example of the case of $\omega_r=2 \omega_z=1$ (more generally
bearing in mind the case of $\omega_r \neq \omega_z$), which will
be considered in our numerical eigenvalue analysis below.
Once again, in this case, 4 modes remain invariant, namely one at
$\omega=0$ (phase invariance), and at $\omega=\omega_z$ and
(double) $\omega=\omega_r$ due to the dipolar oscillations.
The manifold of the modes $(1,0,0)$ and $(0,1,0)$
with frequency $\omega=\omega_r-\omega_z$ at the linear
limit leads to the prediction (using again degenerate perturbation theory)
\begin{eqnarray}
\omega=\left(\omega_r-\omega_z \right) - \frac{1}{3}
\left[\mu - (\omega_r + \frac{3}{2} \omega_z) \right].
\label{eig3}
\end{eqnarray}
On the other hand, the anomalous mode of indices $(0,0,0)$ is
theoretically predicted to move according to
\begin{eqnarray}
\omega=\omega_z - \frac{1}{6}
\left[\mu - (\omega_r + \frac{3}{2} \omega_z) \right].
\label{eig4}
\end{eqnarray}
Then, the sub-manifold of eigenvectors with
$U_1=(\frac{|2,0,0\rangle - |0,2,0\rangle}{\sqrt{2}},0)^T$
and of $U_2=(|1,1,0\rangle,0)^T$ produces two coincident eigenfrequencies (actually two pairs) with
\begin{eqnarray}
\omega=\left(2 \omega_r-\omega_z \right) - \frac{2}{3}
\left(\mu - (\omega_r + \frac{3}{2} \omega_z) \right).
\label{eig5}
\end{eqnarray}

However, we can see that there are additional eigenvalues,
in this case, that acquire comparable values and therefore
manifolds of larger sum $n+m+k$ need to be considered
(up to now we had restricted considerations to $n+m+k=2$).
Among the eigenvalues with $n+m+k=3$, the case of $(0,0,3)$
will have an eigenfrequency of $\omega=1$ in the present setting.
The perturbative calculation in this case yields a prediction
of
\begin{eqnarray}
\omega=2 \omega_z  - \frac{1}{12}
\left[\mu - (\omega_r + \frac{3}{2} \omega_z) \right].
\label{eig6}
\end{eqnarray}
Even more complicated are the degeneracies
occurring at $\omega=1.5$. In addition to
$U_1=(\frac{|2,0,0\rangle - |0,2,0\rangle}{\sqrt{2}},0)^T$
and of $U_2=(|1,1,0\rangle,0)^T$ considered above, there
is the double degeneracy of
$U_1=(\frac{|2,0,0\rangle + |0,2,0\rangle}{\sqrt{2}},0)^T$
and $U_2=(|0,0,4\rangle,0)^T$ and finally that
of $U_1=(|1,0,2\rangle,0)^T$ and $U_2=(|0,1,2\rangle,0)^T$.
The latter, leads to the degenerate eigenfrequencies (again, two pairs)
of the form:
\begin{eqnarray}
\omega=\left(\omega_r+\omega_z \right) - \frac{5}{12}
\left[\mu - (\omega_r + \frac{3}{2} \omega_z) \right].
\label{eig7}
\end{eqnarray}
The former eigenvalues (i.e., the RDS and
the one with $k=4$) are only resonant when
$\omega_r=2 \omega_z$.
In this particular case, we can obtain
the corresponding eigenfrequencies
\begin{eqnarray}
\omega=1.5 - \frac{53 \pm 7 \sqrt{73}}{192} \left(\mu- \frac{7}{4} \right).
\label{eig8}
\end{eqnarray}

We believe it is clear that while the relevant methodology
is entirely general, the logistics of its application vary from case
to case and can be fairly complex. For this reason, we now corroborate
and complement our analysis with detailed numerical computations in the following section.
The above two case studies, of the isotropic regime and of
$\omega_r=2 \omega_z$, will operate as our benchmarks.
We will also
numerically explore other settings such as $\omega_z=2 \omega_r$
as well as, importantly, the spectra of our states of particular focus,
namely the bifurcating (single and double) VRs.

\begin{figure}
\begin{center}
\includegraphics[width=3.3in]{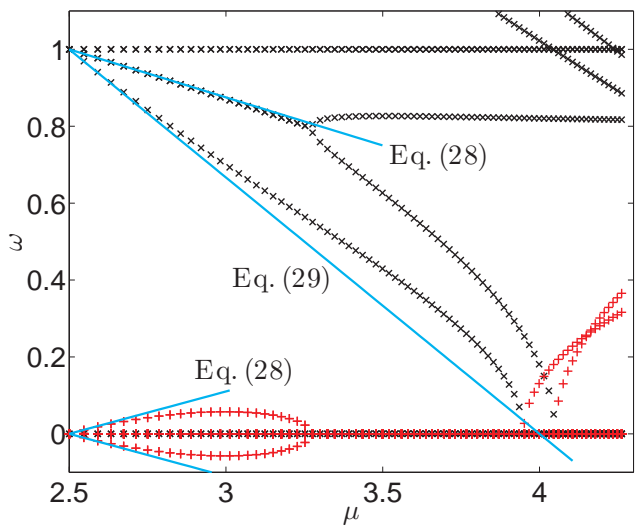}
\caption{(color online)
Spectrum for the dark soliton stationary state for
the isotropic case, $\omega_r=\omega_z$.
Depicted are the eigenfrequencies stemming from the BdG analysis,
as a function of the chemical potential.
Real parts of the eigenfrequencies are denoted by black cross points,
while the imaginary parts of the
eigenfrequencies are denoted by red plus points. The blue solid lines
correspond to theoretical predictions in 
the small-interaction, linear, limit (cf.~text).
}
 \label{Fig:Ene_mu0}
\end{center}
\end{figure}

\section{Numerical Results}\label{sec:numerics}

We identify the stationary solutions by solving the Gross-Pitaevskii equation using a Newton-Krylov scheme~\cite{kelly}. The Bogoliubov linearization spectrum is then obtained by utilizing a spectral basis of noninteracting modes; at least 800 are used for each result herein.
From a numerical standpoint, the problem is reduced to two dimensions by utilizing a Fourier-Hankel method~\cite{ronen} (see also Ref.~\cite{blakie}), which is made possible by the azimuthal symmetry of the trap in the $(x,y)$ plane.

Recall that in Bogoliubov theory every eigenfrequency belongs to a pair, having solutions of opposite sign. From here on, we simplify our discussion by only plotting half of the spectrum, one eigenfrequency from each of these pairs.

\subsection{The dark soliton, $\omega_r=\omega_z$}

\begin{figure}
\begin{center}
\includegraphics[width=3.1in]{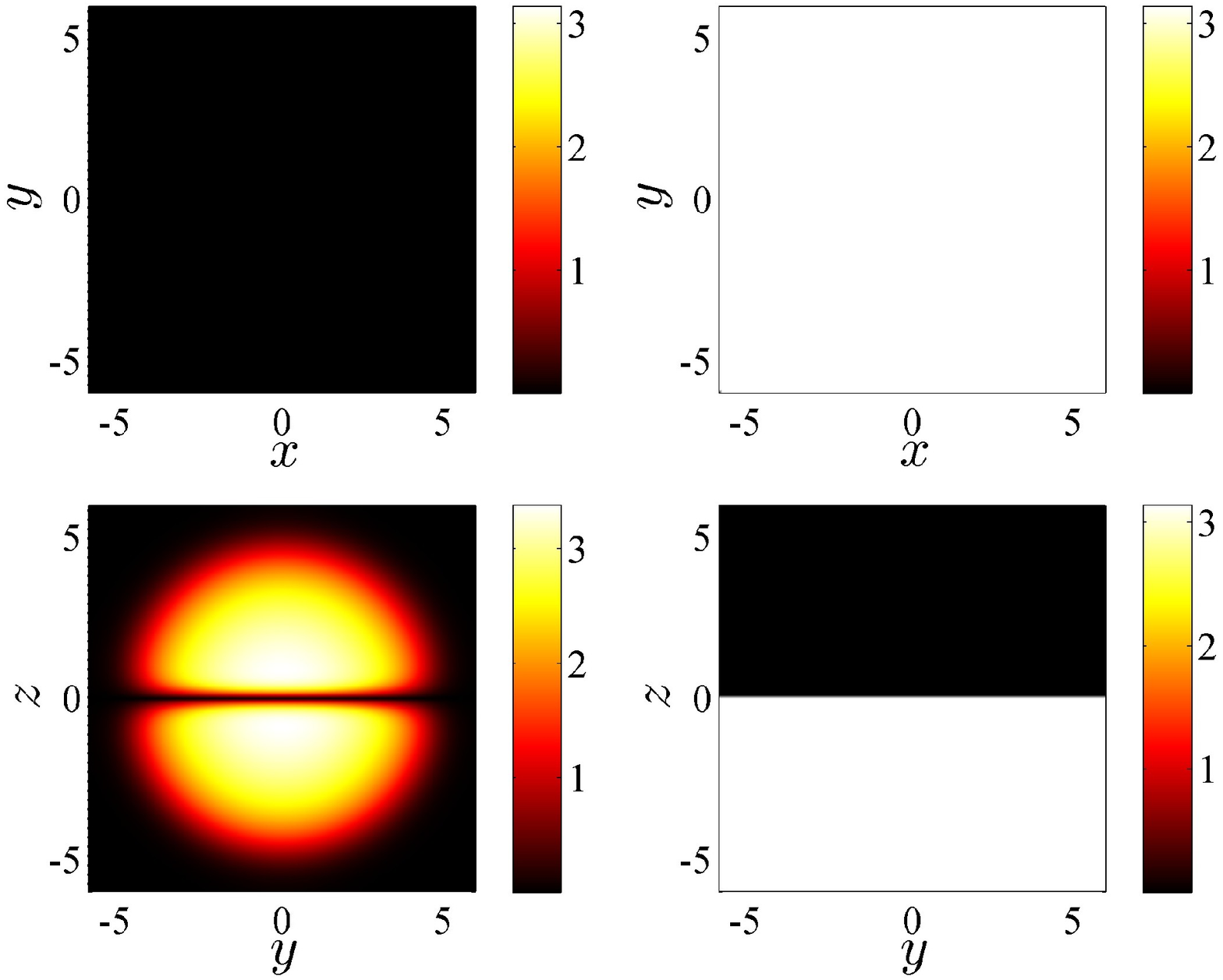}
\includegraphics[width=2.0in]{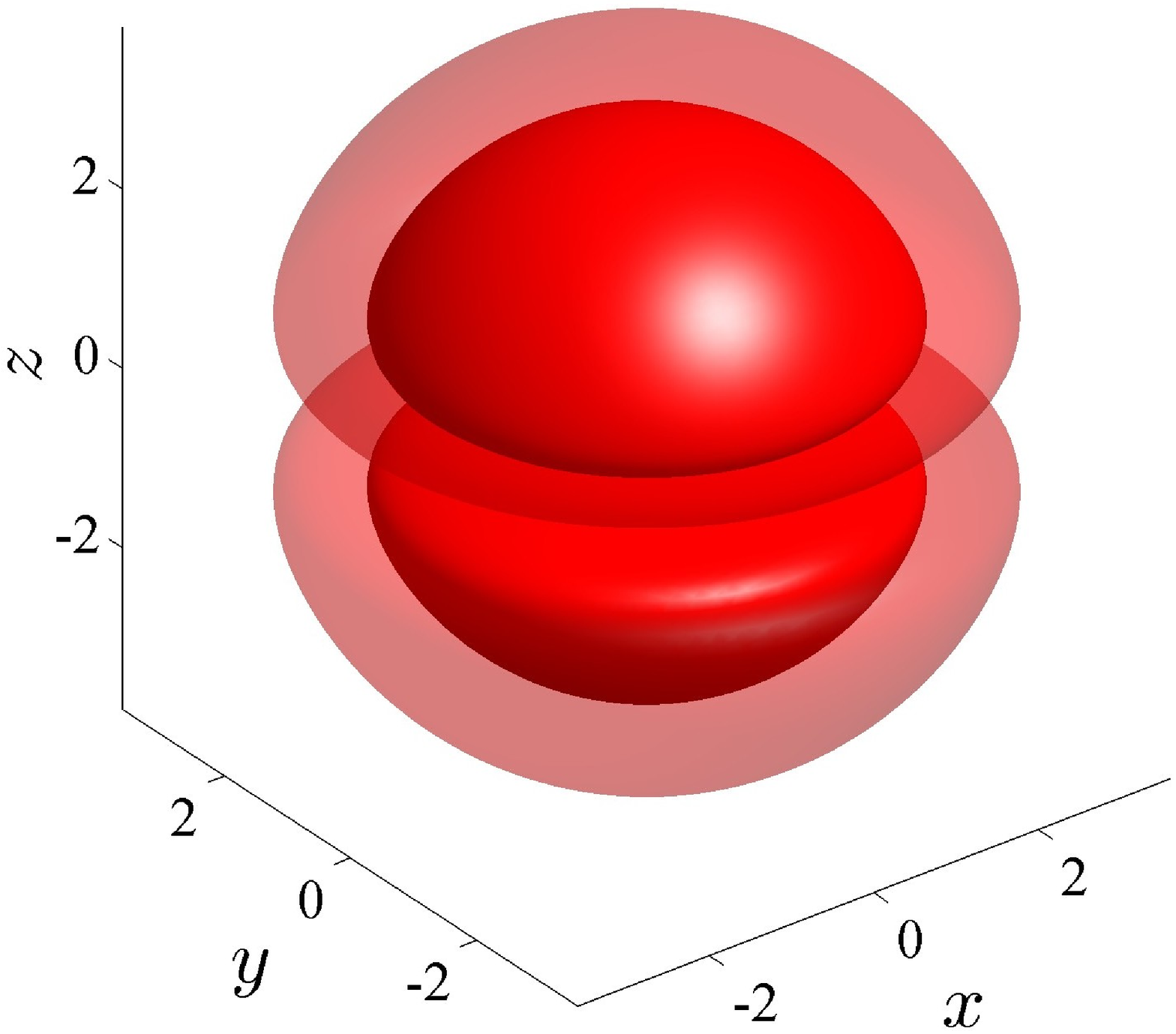}
\caption{(color online)
Dark soliton state for the isotropic case of $\omega_r=\omega_z=1$ and $\mu=12$.
The top four panels show the modulus of the solution (left subpanels) and their
respective argument $\theta$ (right subpanels).
The top row illustrates the $z=0$ plane, where the density vanishes,
and the middle row shows the $x=0$ plane.
The bottom panel shows two isocontour density plots for this state
(the contours correspond to iso-density surfaces at the maximum density
divided by 2.5 and 1.5).
\label{Fig:Sol_wfn}}
\end{center}
\end{figure}

We begin our discussion by considering dark soliton stationary state for the isotropic
case, $\omega_r=\omega_z=1$.
The corresponding results are depicted in Fig.~\ref{Fig:Ene_mu0};
a typical example of the solution itself (both planar cuts,
as well as a full 3D density isocontour plot) is shown in
Fig.~\ref{Fig:Sol_wfn}.
A first observation is that our analytical predictions seem to
agree well 
in the linear limit with the corresponding numerical
results. More specifically, we observe that the complex quartet
of Eq.~(\ref{eig1}) indeed destabilizes the planar dark soliton
in the linear limit, as was originally observed in Ref.~\cite{feder}.
However, as we depart from this limit, the phenomenology reported
in Ref.~\cite{coles} appears to arise, namely the resonant interaction
of the RDS mode and of the anomalous mode ceases.
Thereafter, the anomalous mode maintains a roughly constant frequency,
while the RDS rapidly decreases in frequency, eventually colliding
with the origin (i.e., with $\omega=0$) for a value of $\mu=4.05$ (very
close to the theoretical predictions of $\mu=4$, see Eq.~(\ref{pred2})). Indeed, beyond this critical point, this eigenfrequency becomes imaginary
(equivalently, the eigenvalue becomes real), giving
rise to a symmetry-breaking pitchfork bifurcation. The
daughter branch emerging from this bifurcation (per our analysis),
is the single VR (of charge either $+1$ or $-1$). This will
be further corroborated by the computation of this state below.

Furthermore, we can observe a good agreement also for the two
degenerate eigenfrequencies, stemming from $U_1=(\frac{|2,0,0\rangle - |0,2,0\rangle}{\sqrt{2}},0)^T$
and $U_2=(|1,1,0\rangle,0)^T$ (where $U$ is defined in Eq.~(\ref{Eq:uv}), in accordance with Eq.~(\ref{eig2}).
In this case too, the decrease in frequency eventually leads to
a zero-crossing and a destabilization of the dark soliton
state in favor, in this case, of a double solitonic vortex (2SV) state.
It is important to note that, contrary to the homogeneous (along $z$) case reported in Ref.~\cite{pantof}, here the bifurcation
of the 2SV and that of the single VR do not occur at the same
value of the chemical potential. A consequence of this is that the
dark soliton has already been destabilized by the bifurcation
of the 2SV state when the VR bifurcates, and hence the VR in this
isotropic case is expected to inherit this weak instability close
to that limit.
Note that there are three modes that are invariant at $\omega=0$, one of which is due to the gauge invariance, and two have frequency $\omega_r-\omega_z = 0$, reflecting the rotational invariance of the planar dark soliton.


\begin{figure}
\begin{center}
\includegraphics[width=3.3in]{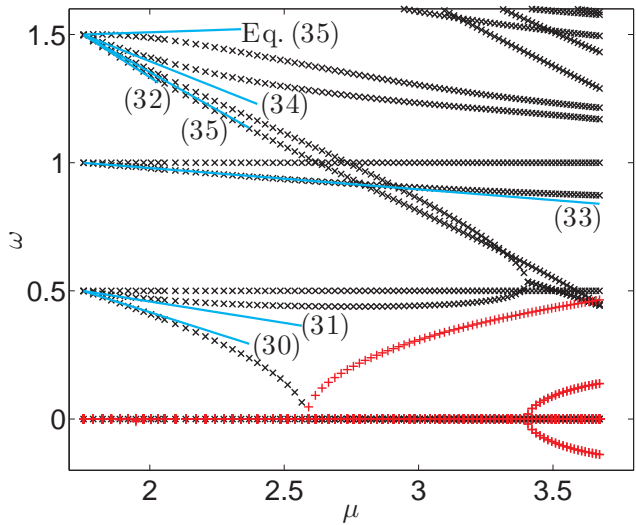}
\caption{(color online)
Spectrum for the dark soliton stationary state for $\omega_r=2 \omega_z=1$.
Points have the same meaning as in Fig.~\ref{Fig:Ene_mu0}.
}
\label{Fig:Ene_mu1}
\end{center}
\end{figure}

\subsection{The dark soliton, $\omega_r=2\omega_z$}

We now turn to the investigation of our second analytically examined
case, namely of $\omega_r=2 \omega_z=1$.
The comparison between the analytical prediction and the numerical
results for the stability spectrum is depicted in Fig.~\ref{Fig:Ene_mu1}.
Remarkably, we can see in this
case too ---despite the considerable complexity of the bifurcation diagram---
the high accuracy of our eigenvalue predictions near 
the linear limit. While in this case there is no quartet
from the linear limit, the instability emerges due to the rapid
decrease of the modes (accurately predicted by Eq.~(\ref{eig3}), near the linear limit)
pertaining to the manifold of $(1,0,0)$ and $(0,1,0)$. We remark that, in the isotropic case, this pair of modes was equi-energetic
with the 
dark soliton, allowing the construction from the linear limit of the single solitonic vortex.
Here, however, it is at $\mu = 2.58$ that the relevant instability emerges, giving rise to the solitonic vortex state. Notably, this also agrees with the analytical prediction of $\mu=2.5$, see Eq.~(\ref{pred6}).
The only other instability that can be observed is given by the
collision of the RDS,
predicted by Eq.~(\ref{eig8}), with the anomalous mode of
Eq.~(\ref{eig4}). This, once again, leads to a quartet
of eigenfrequencies and only for considerably larger values
of $\mu$ (not included in the figure) will the VR bifurcate.
Additional predictions such as the manifold of
$U_1=(\frac{|2,0,0\rangle - |0,2,0\rangle}{\sqrt{2}},0)^T$
and of $U_2=(|1,1,0\rangle,0)^T$ through Eq.~(\ref{eig5}),
as well as the higher index ($n+m+k$) of Eqs.~(\ref{eig6})
and~(\ref{eig7}) are also reasonably accurately captured.

\begin{figure}
\begin{center}
\includegraphics[width=3.3in]{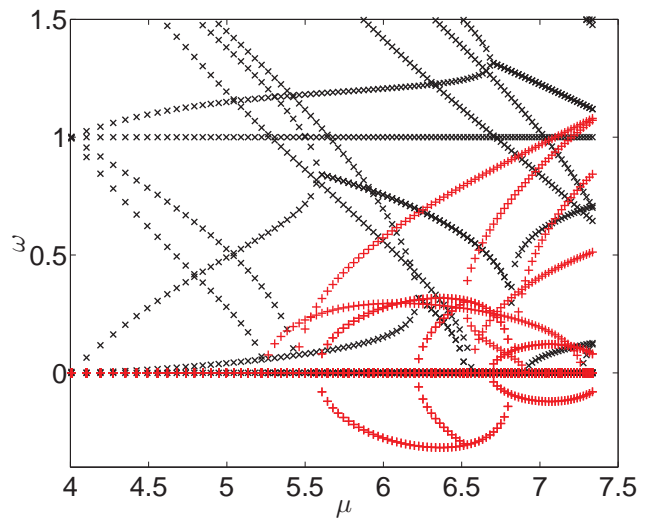}
\caption{(color online)
Spectrum for the dark soliton stationary state for $\omega_r=\omega_z/2=1$.  Otherwise, similar to Fig.~\ref{Fig:Ene_mu0}.
Here the 
ring dark soliton and the planar dark soliton have the same energy,
and hence the VR
can bifurcate immediately from the linear limit. The principal
instabilities
of the dark soliton in this case arise due to higher modes
(with $n+m=3$), although lower-order modes contribute to
oscillatory instabilities associated with complex eigenfrequency quartets.
 \label{Fig:Ene_mu2}}
\end{center}
\end{figure}

\subsection{The dark soliton, $\omega_z = 2\omega_r$}

Finally, we also touch upon the case of $\omega_z=2 \omega_r=2$ 
---cf.~Fig.~\ref{Fig:Ene_mu2}. In this case, the situation is considerably
more complex and numerous instabilities appear to arise.
Nevertheless, in this case too, our understanding of the linear limit
may provide a reasonable set of guidelines for understanding
the relevant phenomenology. In this case, we observe
four instabilities associated with imaginary eigenfrequencies
and three associated with eigenfrequency quartets.
Among the latter, the first to arise is the degenerate pair (the cross states) of
eigenfrequencies, emerging from the origin with negative energy and colliding with a higher-order (in $n+m+k$) mode at $\mu=5.59$.
The second quartet begins at $\mu=6.21$ and involves the RDS (arising from the origin with negative energy) colliding with a higher energy mode. It should be remarked that the presence of the RDS at the linear limit enables
(as discussed earlier) the bifurcation of the VR already at the linear limit.
This, in turn, implies that the VR may be expected
to be robust in this case (however, see the detailed results below).
Finally, the third
collision involves the degenerate modes $(1,0,0)$ and $(0,1,0)$ which
are also now anomalous due to their negative energy and
are growing from the linear limit. This pair collides
with higher order modes and a quartet develops for $\mu>6.69$.
The first two imaginary instabilities arising
are, surprisingly, {\it not} caused
by lower-order modes, but rather from the higher-order ones
$(3,0,0)$ and $(0,3,0)$, as well as from $(1,2,0)$ and
$(2,1,0)$. Both of these sets of eigenfrequency pairs decrease,
as we depart from the linear limit, leading to an imaginary eigenfrequency instabilities for $\mu>5.24$ and $\mu>5.44$, respectively.
Hence, we see that in this case too, we can form a qualitative
picture of the stability landscape merely by knowing and suitably
appreciating the linear eigenfrequency/eigenmode picture.

\begin{figure}
\begin{center}
\includegraphics[width=3.3in]{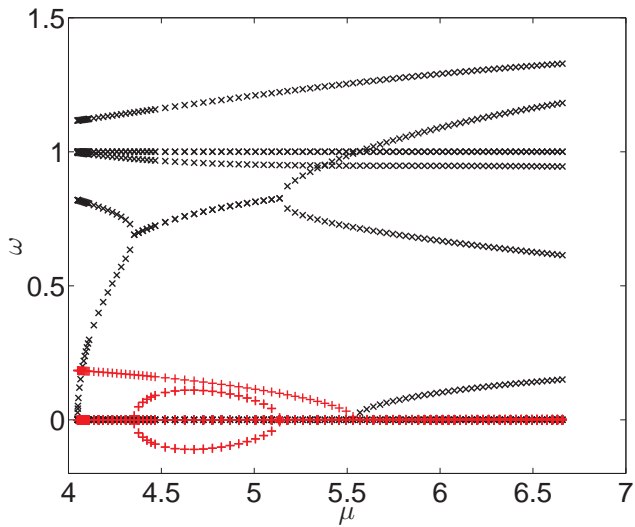}
\caption{(color online)
Single VR for $\omega_r=\omega_z$. Similar to that of Fig.~\ref{Fig:Ene_mu0},
the present figure shows the BdG spectrum
(this spectrum coincides with that of the dark soliton when the
VR bifurcates from the it at $\mu = 4.05$;
cf.~Fig.~\ref{Fig:Ene_mu0}). Notice that despite oscillatory and
imaginary eigenfrequency-related instabilities (discussed in the
text) for small values of $\mu$,
the VR is stabilized in the large chemical potential limit.
 \label{Fig:Ene_mu3}}
\end{center}
\end{figure}

\begin{figure}
\begin{center}
\includegraphics[width=3.1in]{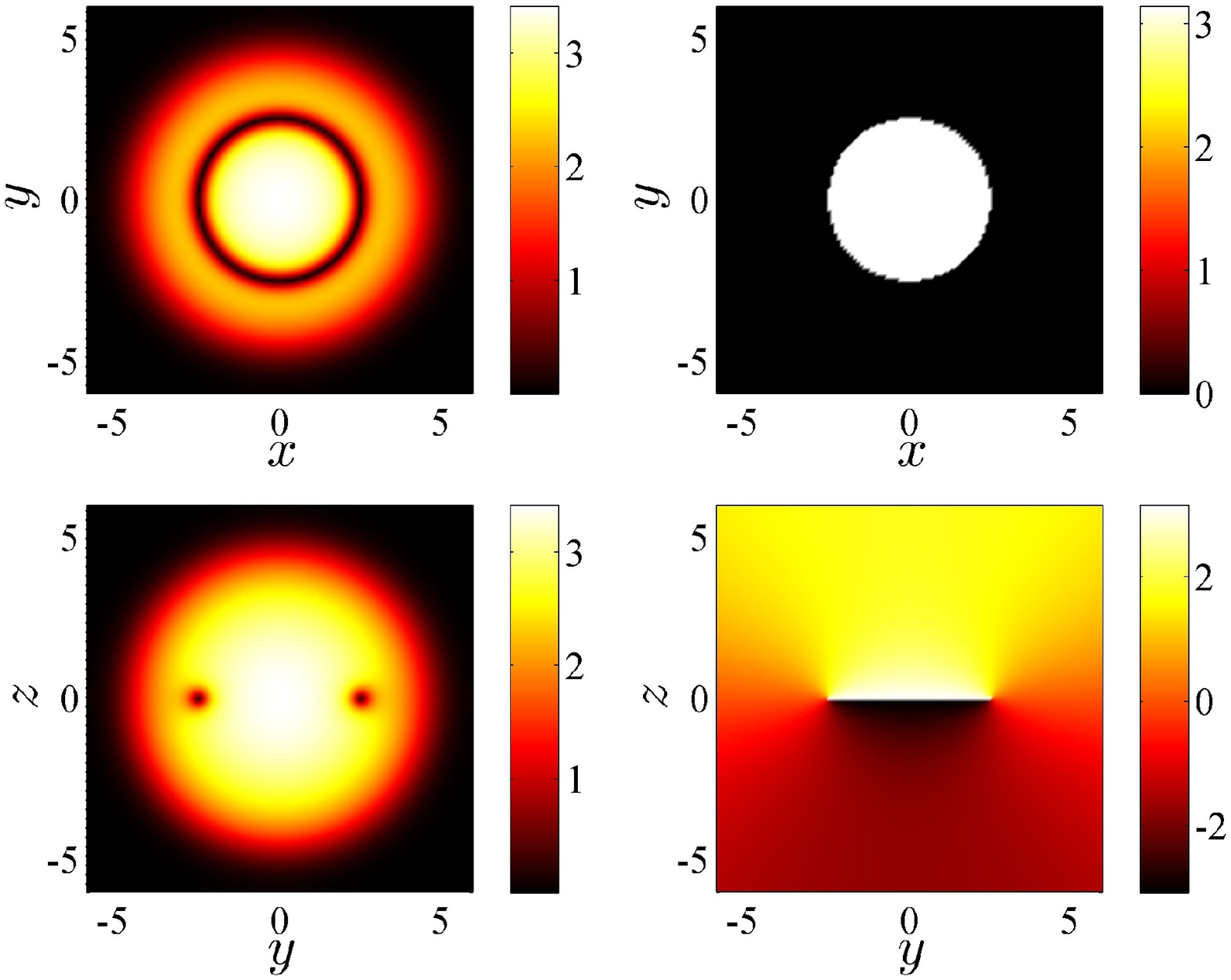}
\includegraphics[width=2.0in]{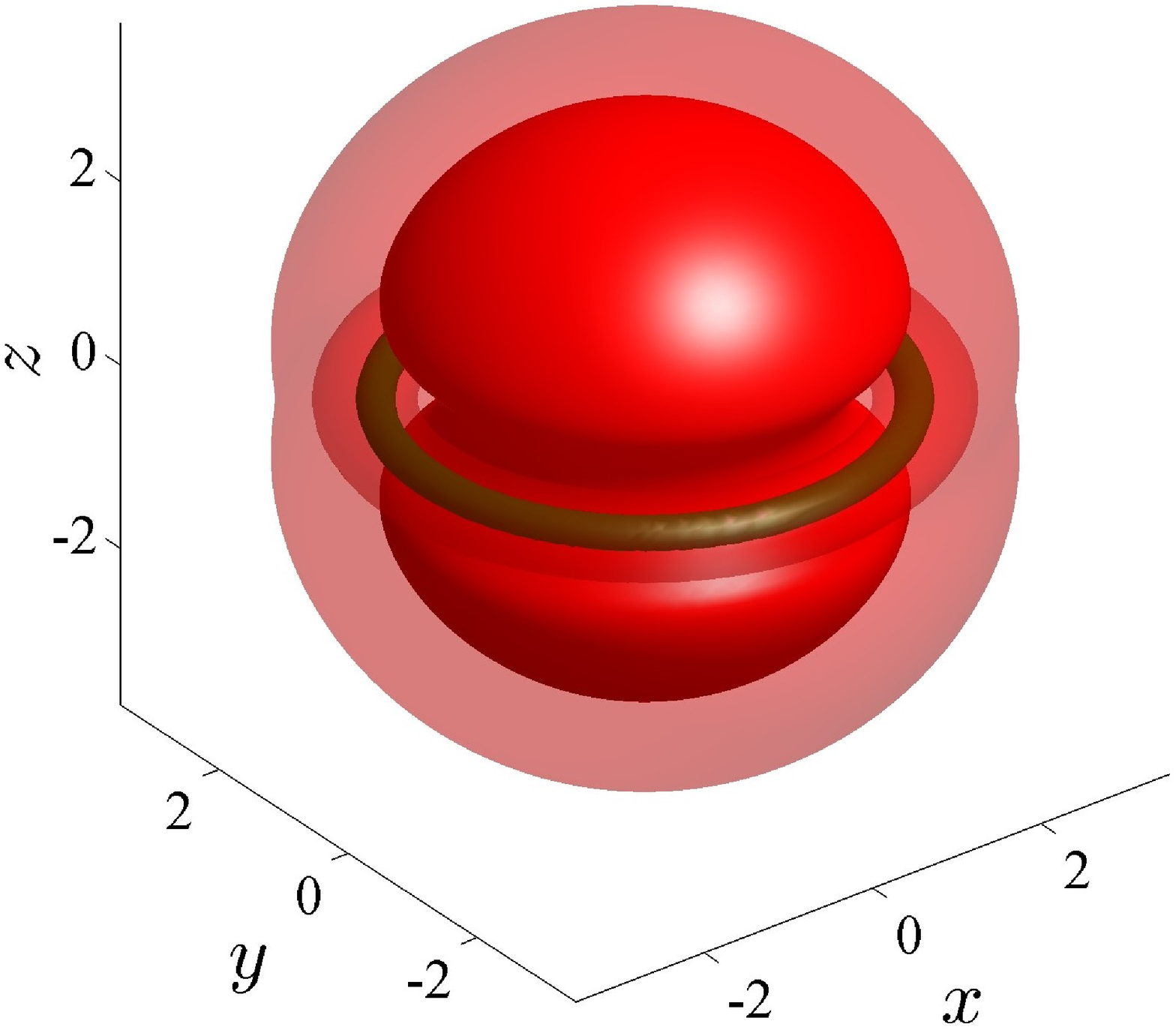}
\includegraphics[width=3.1in]{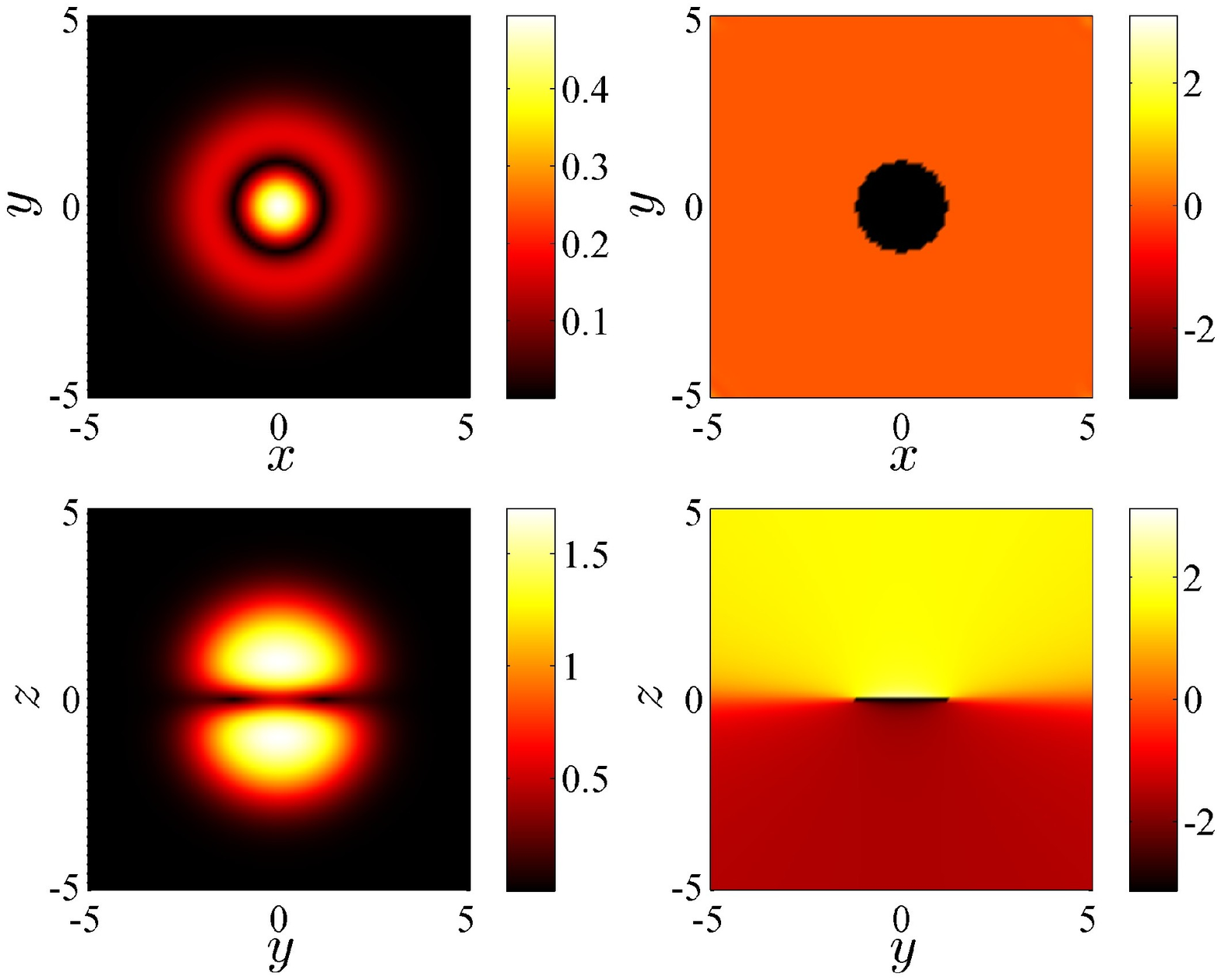}
\caption{(color online)
Single VR state for $\omega_r=\omega_z=1$.
Similar layout as in Fig.~\ref{Fig:Sol_wfn}.
The cross sections of the top two rows and the density isocontour
of the middle panel show a prototypical example of the bifurcating
branch (for $\mu > 4.05$) of the single VR in the large chemical
potential limit (here at $\mu=12$), where the relevant state has been
stabilized against small-amplitude perturbations. The darker (green)
isocontour corresponds to an iso-density at the maximum density divided by 8
plotted inside the bulk of the cloud.
The bottom two rows correspond to a VR solution
right after its bifurcation at $\mu=4.05$, clearly illustrating
its ring dark soliton character in the $(x,y)$ plane and
its ``emerging vorticity'' in the phase which is reminiscent of
a vortex dipole (despite the resemblance with a dark soliton
in the density).
\label{Fig:Sol_vr1}}
\end{center}
\end{figure}

\subsection{The single vortex ring, $\omega_r=\omega_z$}

We now turn our attention to the case of the VR,
first considering the isotropic trap; see Fig.~\ref{Fig:Ene_mu3}
for the BdG spectrum and Fig.~\ref{Fig:Sol_vr1} for
an illustration of the relevant state far from (top two rows) and
near to (bottom two rows) its bifurcation point.
As expected, the eigenfrequency pertaining to the RDS
$U_1=(\frac{|2,0,0\rangle + |0,2,0\rangle}{\sqrt{2}},0)^T$, grows along the real axis.
Recall that, for the dark-soliton stationary state, the RDS was the mode that became unstable and gave rise to the VR.
However, for larger values of the chemical potential
(i.e., for $\mu \in [4.35,5.14]$), and in a way reminiscent
to what happens for the dark soliton in the vicinity of
the linear limit, the RDS collides with the anomalous, negative energy
mode $U_2=(0,|0,0,0\rangle)^T$,
giving rise to an oscillatory instability (and an associated
eigenfrequency quartet). This instability disappears
for larger values of $\mu > 5.14$.

Importantly, since
the VR bifurcated from the dark soliton state (both with $\omega_r=\omega_z$) then at the point of bifurcation their spectra should be identical, cf.~Figs.~\ref{Fig:Ene_mu0} and~\ref{Fig:Ene_mu3} at $\mu = 4.05$.
Recall, though, that at the bifurcation point another instability was already present in the spectrum of the dark soliton due to the prior bifurcation of the 2SV. Consequently, the VR is endowed with this instability from birth.
However, upon increase of the chemical potential, we observe that the eigenfrequency associated with
this inherited instability
(corresponding to a doubly degenerate
subspace
associated with $U_1=(\frac{|2,0,0\rangle - |0,2,0\rangle}{\sqrt{2}},0)^T$
and of $U_2=(|1,1,0\rangle,0)^T$, i.e., the ``cross states'')
decreases and eventually
leads to a complete stabilization of the VR for $\mu>5.54$.
It is due to this stabilization that
such VR patterns can be observed
robustly in the large chemical potential (Thomas-Fermi) limit.
In the latter limit, the VR acquires particle-like characteristics;
see, e.g., Ref.~\cite{horng} for the dynamics of a single VR inside
a trap and Ref.~\cite{caplan} for the interactions between multiple
VRs. The combination of these features and the connection
of the spectral features of the VR with the above particle-like
characteristics will be explored separately in a future publication.

\begin{figure}
\begin{center}
\includegraphics[width=3.3in]{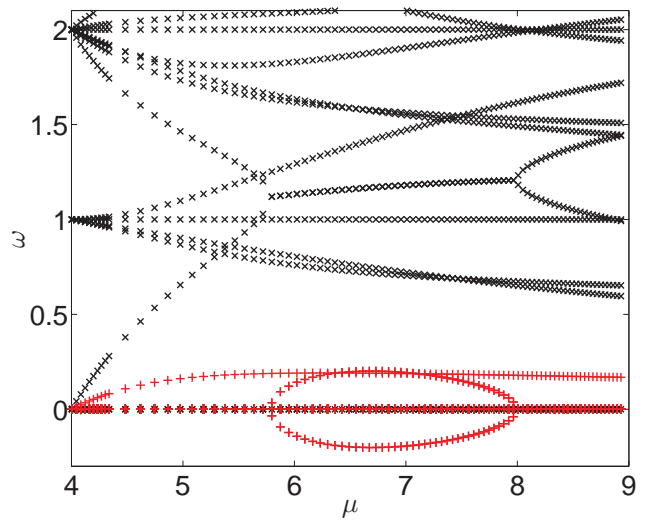}
\caption{(color online) Spectrum for the single VR when $\omega_r=\omega_z/2=1$,
similar to Fig.~\ref{Fig:Ene_mu3}.
Here, the VR emerges from the linear limit.}
\label{Fig:Ene_mu4}
\end{center}
\end{figure}

\begin{figure}
\begin{center}
\includegraphics[width=3.3in]{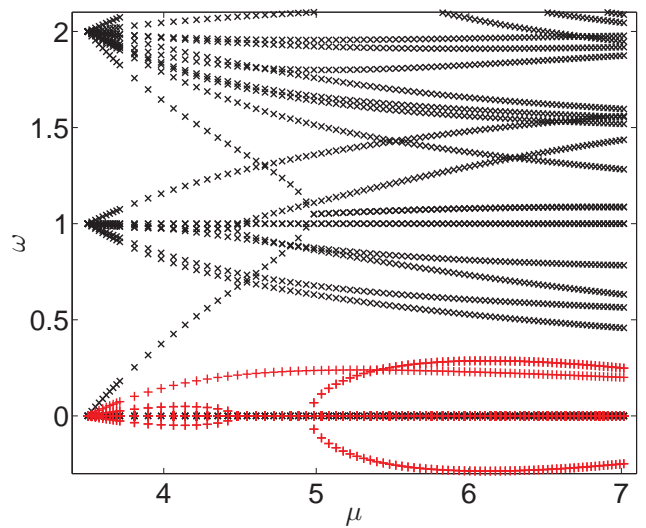}
\caption{(color online)
Spectrum for the double vortex ring (2VR) for the
isotropic case $\omega_r=\omega_z=1$.
Points have the same meaning as previous figures.
 \label{Fig:Ene_mu5}}
\end{center}
\end{figure}

\subsection{The single vortex ring, $\omega_z=2\omega_r$}

We now briefly examine the VR for the special case where it bifurcates from
the linear limit, namely when $\omega_z=2 \omega_r$ (and hence
the energies of the dark soliton and the RDS become degenerate,
enabling the construction of a VR already at the linear limit,
in accordance with Eq.~(\ref{VR})).
The corresponding BdG spectrum is depicted in Fig.~\ref{Fig:Ene_mu4}.
Interestingly, here, we observe that the VR is, in fact, unstable
immediately (i.e., as soon as we depart from the linear limit). This instability
is manifested through a degenerate imaginary pair
of eigenfrequencies, again associated with the cross states
$U_1=(\frac{|2,0,0\rangle - |0,2,0\rangle}{\sqrt{2}},0)^T$
and of $U_2=(|1,1,0\rangle,0)^T$.
In this limit, the latter states {\it also} bifurcate from the
linear limit and apparently have lower energy than the VR. We note, however, that as we proceed
to larger values of the chemical potential, the
instability appears to (weakly) decrease
and, hence, VRs may again be long-lived in
the Thomas-Fermi limit of large $\mu$. Additionally, it is worthwhile
to note that once again the collision
of $U_1=(\frac{|2,0,0\rangle + |0,2,0\rangle}{\sqrt{2}},0)^T$ with the anomalous mode $U_2=(0,|0,0,0\rangle)^T$ leads to a resonance
and an oscillatory instability for the interval $5.78 \leq \mu \leq 8.00$.

We also remark that, for this stationary state as well, the relevant eigenvalue
count can be performed.
Consider the eigenvalues of the
operator ${\cal L}-\mu_0$ (with $\mu_0=4\omega_r$) to be $\omega_r (n + m + 2 k) - 2 \omega_r$.
We find that there are
(a) 4 modes at the spectral plane origin (3 climb from the origin with increasing $\mu$, as discussed above, and one remains at the origin due to gauge invariance), there are
(b) 8 modes at $\omega=\omega_r$, two of which are anomalous, while there are
(c) 10 modes at $\omega=2 \omega_r$, one of which is anomalous
and gives rise to the oscillatory instability discussed above.
Hence, given the technical complications of considering the relevant
perturbative analysis (and also the qualitative understanding afforded
to us by means of the above analysis), we will not pursue this further
here.

\begin{figure}
\begin{center}
\includegraphics[width=3.1in]{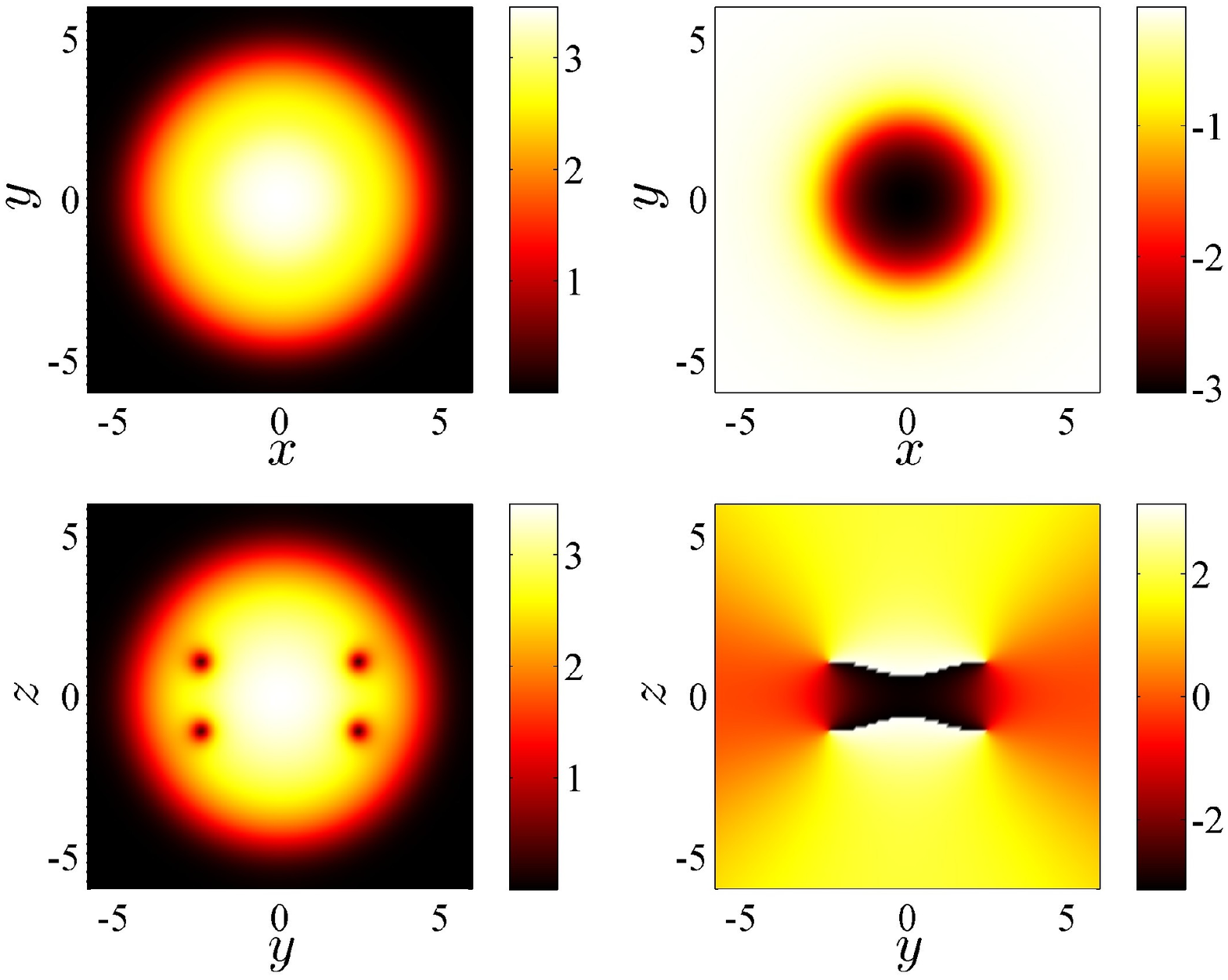}
\includegraphics[width=2.0in]{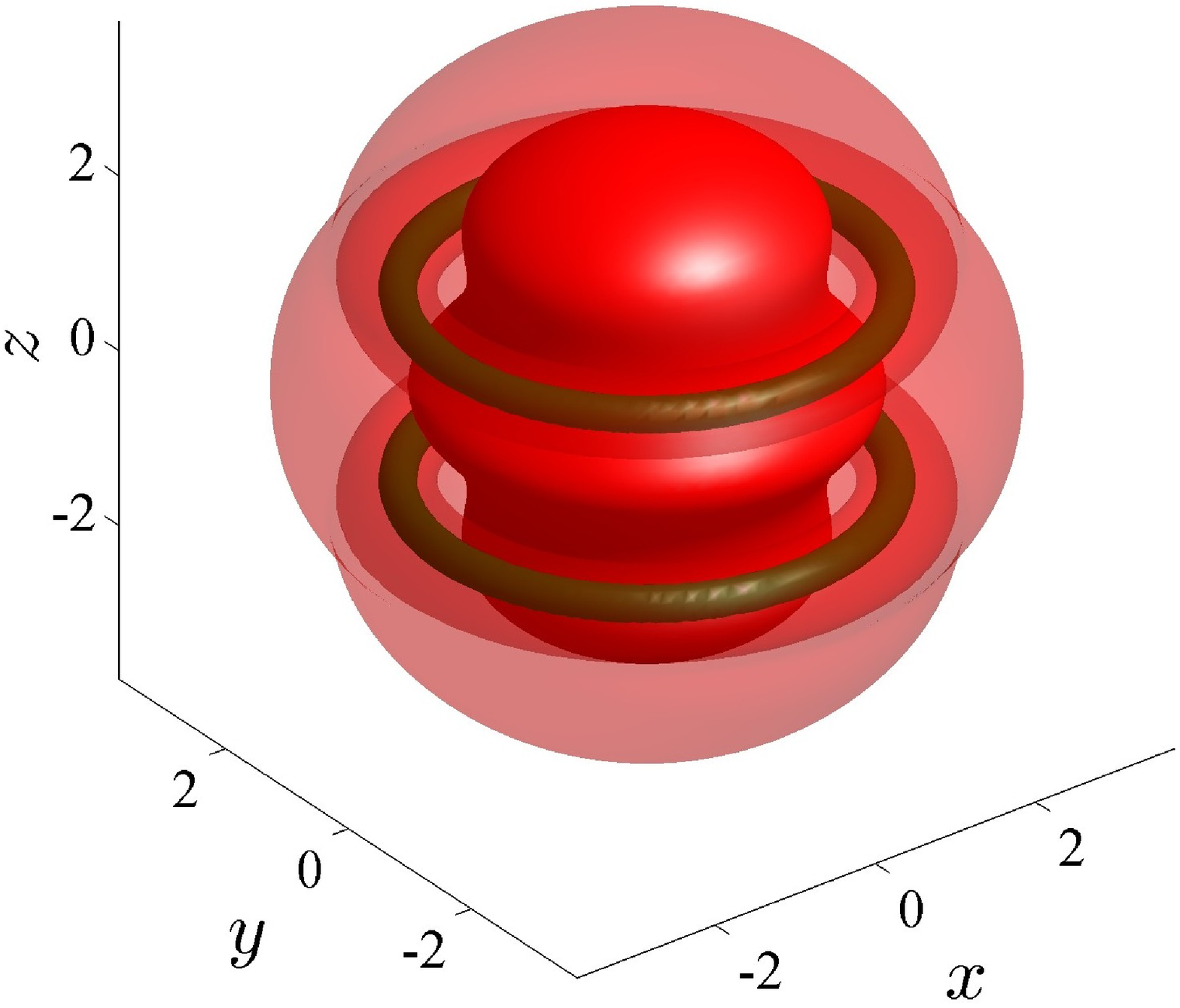}
\caption{(color online)
Double VR (2VR) state for the isotropic case $\omega_r=\omega_z=1$ and $\mu=12$.
Similar layout as in Figs.~\ref{Fig:Sol_wfn} and \ref{Fig:Sol_vr1}.
This state bifurcates from the linear limit at $\mu=7 \omega_r/2$.
\label{Fig:Sol_vr2}}
\end{center}
\end{figure}

\subsection{The double vortex ring, $\omega_z=\omega_r$}

Finally, we also examine the double-vortex ring (2VR) state,
bifurcating from the linear limit of $\mu_0=7 \omega_r/2$
in the isotropic case of $\omega_r=\omega_z$,
see Figs.~\ref{Fig:Ene_mu5} and \ref{Fig:Sol_vr2}.
Here the situation is
even more complicated, with 6 degenerate eigenfrequencies at $\omega=0$,
13 modes (3 anomalous ones) at $\omega=\omega_r$,
and so on. Hence, we will again restrict our considerations to some
qualitative remarks. We note that the state will be {\it immediately}
unstable in this isotropic limit, due to an imaginary eigenfrequency pair (the cross states), bifurcating
from $0$ as soon as the 2VR state emerges. Moreover, the resonance
of the degenerate anomalous pair at $\omega=\omega_r$ with the corresponding
positive (same) energy modes leads to a scenario again reminiscent
of Ref.~\cite{coles} in that an oscillatory instability emerges
from the linear limit, although it is terminated at $\mu=4.47$.
An additional collision of a mode bifurcating from the origin with
the anomalous mode $U_2=(0,|0,0,0\rangle)^T$ leads to an additional
eigenfrequency quartet for $\mu>4.96$. A key observation here is
that these instabilities (both the imaginary one and the oscillatory one for
$\mu>4.96$) were found to persist throughout the interval of
parameters considered herein. However, for large values of
$\mu$, we observe a weak tendency for both of these instabilities
towards decreasing growth rates. The latter feature appears to
be qualitatively consonant with the observations of Ref.~\cite{ionut},
suggesting that, for intermediate parameters, 2VR states
were less robust than for large values of the corresponding
nonlinearity-controlling parameter. Once again, the theme of
the Thomas-Fermi limit will be the subject of separate work,
focusing on the dynamics and interactions of VRs as particle
entities.

\section{Conclusions and Future Challenges}
\label{sec:conclu}

In conclusion, we have systematically examined the full
eigenvalue spectrum by linearization around a dark soliton in a 3D setting,
building on considerable earlier work, most notably
that of Ref.~\cite{feder}.
Extending that work, we have shown how to predict not only the
formation and bifurcation of the (single) vortex ring state, but also
of other states,
such as the double-vortex-ring,
the single- and the double-solitonic vortices, among others.
We have explained how the emergence of
these states can be predicted on the basis of a Galerkin-type, two-mode, approach. This is analogous
to how a vortex pair, and more complex states involving
multiple vortices, were predicted to arise in 2D 
(see, e.g., Refs.~\cite{middelphysd,middel2}), a step that
subsequently led to their experimental realization~\cite{hallpra}.
We have explored the conditions necessary for the vortex ring to emerge through a bifurcation from
the planar dark soliton state,
and explained when it can arise from the linear limit (when the
energy of the ring dark soliton and that of the dark soliton coincide). Conversely, the vortex ring may arise through a bifurcation from the
ring dark soliton, when the latter possesses lower energy than the
planar dark soliton state. We have also highlighted similarities and
differences from important related work such as the contribution
of Ref.~\cite{pantof}.

In addition to giving a qualitative
characterization of the relevant spectra and bifurcations,
in the vicinity of the linear limit, we also employed perturbation
theory (often in its more complex degenerate form) in order to
quantitatively characterize the eigenvalues responsible for the
relevant instabilities and the emergence of new branches.

Upon obtaining a systematic understanding of when the vortex
rings arise, we turned our attention
to their spectral stability properties, by numerically solving the Bogoliubov-de
Gennes equations.
We thus found
that at the isotropic limit, the single vortex ring may be
unstable ``at birth'', but can be stabilized at higher chemical
potentials.
In the anisotropic case, where the vortex ring
state emerges from the linear limit, it is also immediately unstable
but this instability can be relatively weak in different parametric
regimes, such as that of large chemical potential. 
Similar features were found for the double vortex ring, in qualitative
agreement with earlier dynamical observations~\cite{ionut}.

We believe that this study paves the way for a deeper
understanding of such vortex ring states, offering
an unprecedented
view of
their spectral features. 
 In this light, there are numerous aspects worthy of further
exploration.
A more technical 
example involves the attempt to systematically
characterize the spectrum of multiple vortex rings in the special cases where
they emerge in the linear limit, i.e., at $\omega_z=2 \omega_r/k$
for the $k$-th vortex ring state. A more intriguing aspect
for our considerations is to explore the opposite limit
more systematically, namely the Thomas-Fermi realm, where
the single and multiple vortex rings possess particle
characteristics. While in this area there is some work already
done, it is rather incomplete. 
For instance, in Ref.~\cite{horng}, a theoretical approximation
of the vortex ring internal mode spectrum was obtained at this
``particle limit'', but it was never tested against the full
Bogoliubov-de Gennes spectrum or the three-dimensional GPE.
On the other hand, in Ref.~\cite{caplan}, a particle picture is
derived and favorably compared to the GPE results, but this
is only done in the absence of a trap. Obtaining a conclusive
spectral picture for single and multi-vortex rings
in the presence of a trap
would certainly be of paramount importance for our understanding
of the particle-like character of these complex states and of
their interactions within trapped BECs.
Finally, once these aspects are addressed, one can think of
not only examining the role of thermal and/or quantum fluctuations
on these rings, but also importantly of generalizing them
in multi-component systems, where elaborate spinorial
states exist in the form of skyrmions~\cite{ruost} and
monopoles~\cite{simula}, among many others.
Such studies are currently in progress and will be reported
in future publications.

\begin{acknowledgments}

R.N.B.~would like to thank D.~Baillie and R.M.~Wilson for useful discussions.
W.W.~acknowledges support from NSF (grant No.~DMR-1208046).
P.G.K.~gratefully acknowledges the support of
NSF-DMS-1312856, as well as from
the US-AFOSR under grant FA950-12-1-0332,
and the ERC under FP7, Marie Curie Actions, People,
International Research Staff Exchange Scheme (IRSES-605096) and insightful
discussions with Prof. Ionut Danaila.
P.G.K.'s work at Los Alamos is supported in part by the U.S. Department of Energy.
R.C.G.~gratefully acknowledges the support of NSF-DMS-1309035.
The work of D.J.F.~was partially supported by the Special Account
for Research Grants of the University of Athens.
This work was performed under the auspices of the Los Alamos National
Laboratory, which is operated by LANS, LLC for the NNSA of the U.S.~DOE
under Contract No.~DE-AC52-06NA25396.

\end{acknowledgments}


\end{document}